\documentclass{article}
\usepackage{spconf,amsmath,graphicx}

\usepackage{enumitem}
\setlist{nosep, leftmargin=14pt}

\usepackage{mwe} 
\usepackage{algorithm,algpseudocode}
\usepackage{booktabs}

\def\L{{\cal L}}

\newcommand{\eg}{{\it e.g.}}
\newcommand{\ie}{{\it i.e.}}
\newcommand{\argmin}{\mathop{\mathrm{argmin}}\limits}

\title{Evidence-empowered Transfer Learning for Alzheimer's Disease}

\name{%
  \begin{tabular}{c}
  {\it Kai Tzu-iunn Ong}$^{1}$\qquad {\it Hana Kim}$^{2}$\qquad {\it Minjin Kim}$^{1}$\qquad {\it Jinseong Jang}$^{3}$\qquad {\it Beomseok Sohn}$^{4}$ \\
  {\it Yoon Seong Choi}$^{4}$\qquad {\it Dosik Hwang}$^{3}$\qquad {\it Seong Jae Hwang}$^{1}$\qquad {\it Jinyoung Yeo}$^{1}$\sthanks{Corresponding author (Email: jinyeo@yonsei.ac.kr)}
  \end{tabular}%
}

\address{Department of Artificial Intelligence$^{1}$, Computer Science$^{2}$, Electrical Engineering$^{3}$, Yonsei University\\Department of Radiology$^{4}$, College of Medicine, Yonsei University}

\begin{document}

\maketitle

\begin{abstract}

Transfer learning has been widely utilized to mitigate the data scarcity problem in the field of Alzheimer's disease (AD). 
Conventional transfer learning relies on re-using models trained on AD-irrelevant tasks such as natural image classification. However, it often leads to negative transfer due to the discrepancy between the non-medical source and target medical domains. To address this, we present \textit{evidence-empowered transfer learning} for AD diagnosis. Unlike conventional approaches, we leverage an AD-relevant auxiliary task, namely \textit{morphological change prediction}, without requiring additional MRI data. In this auxiliary task, the diagnosis model learns the evidential and transferable knowledge from morphological features in MRI scans. 
Experimental results demonstrate that our framework is not only effective in improving detection performance regardless of model capacity, but also more data-efficient and faithful. 

\end{abstract}


\section{Introduction} \label{sec:intro}

Recently, machine learning has gained much attention in addressing the accurate diagnosis of Alzheimer's disease (AD). Despite its effectiveness, high-capacity models such as convolutional neural networks (CNNs) still suffer from the lack of training data due to the limited availability of public image data for AD~\cite{setio2017validation, simpson2019large}. A popular solution to the data scarcity problem is transfer learning with arbitrary auxiliary tasks, \eg, re-using models trained on \emph{natural image classification}, tuning them on the target task. However, we observe that such a naive remedy leads to negative transfer to AD diagnosis.

\begin{figure}[ht]
    \begin{minipage}[t]{1.0\linewidth}
        \centering
        \includegraphics[width=\textwidth]{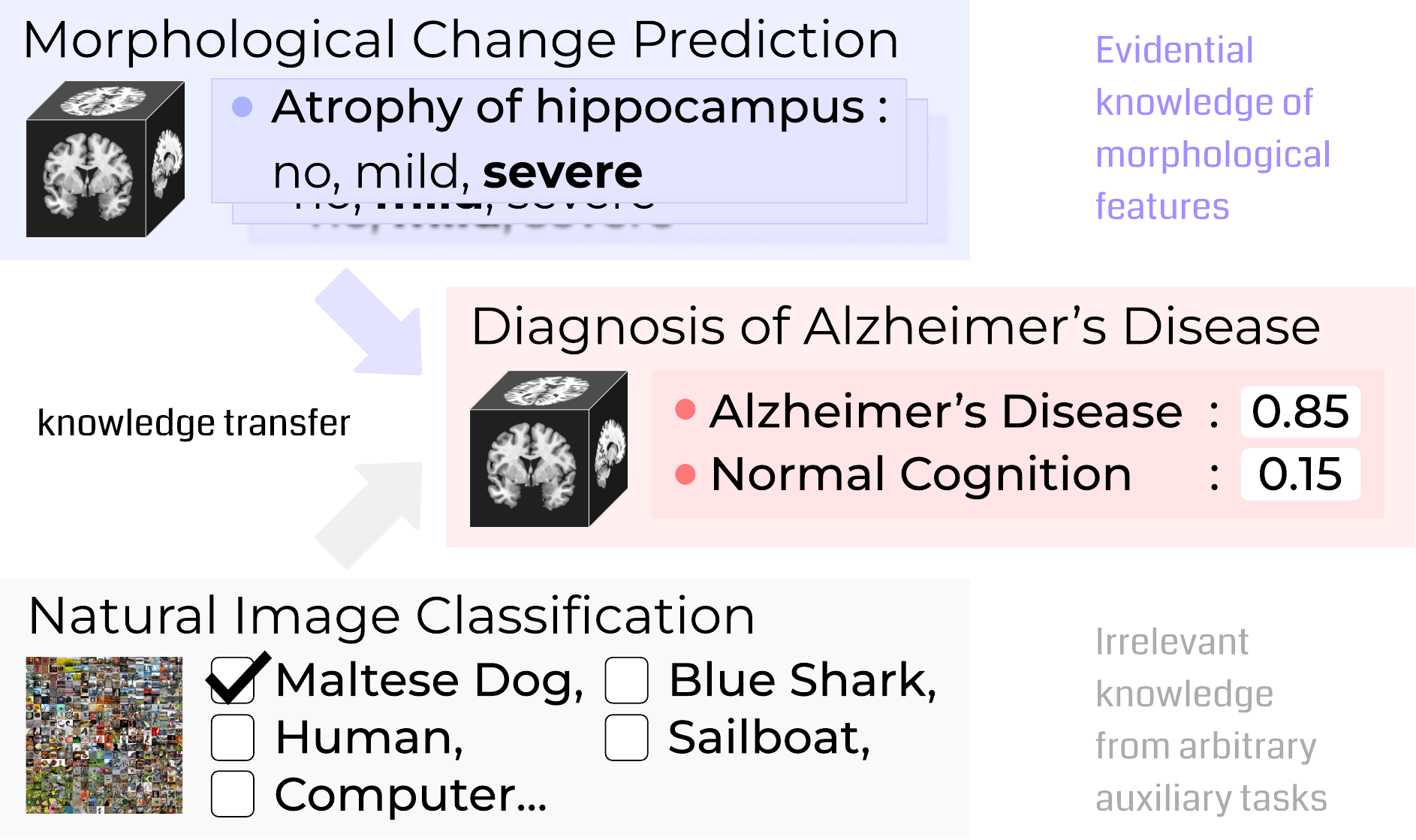}
    \end{minipage}
    \caption{\textbf{Illustration of transfer learning for AD diagnosis:} The proposed framework vs. conventional transfer learning.}
    \label{fig:1}
\end{figure}

In this paper, we present a simple yet effective transfer learning with an AD-relevant auxiliary task named \emph{morphological change prediction}, which leverages the morphological features in 3D MRI brain images. The conventional way to use the morphological features is to extract summary measures (\eg, subcortical volume and cortical thickness measurements) and use them as additional input features~\cite{kruthika2019multistage}, but it is challenging to effectively model such multimodal information (\ie, 3D image and summary measures). 
By contrast, as illustrated in Figure~\ref{fig:1}, we focus on learning the transferable knowledge where the summary measures are discretized into the categorical severity of morphological changes (\eg, \texttt{no}, \texttt{mild}, and \texttt{severe} classes of cortical atrophy, ventricle area enlargement, and hippocampal volume shrinkage) as ground-truth labels of the designed auxiliary task.

\begin{figure*}[ht]
    \centering
    \includegraphics[width=\textwidth]{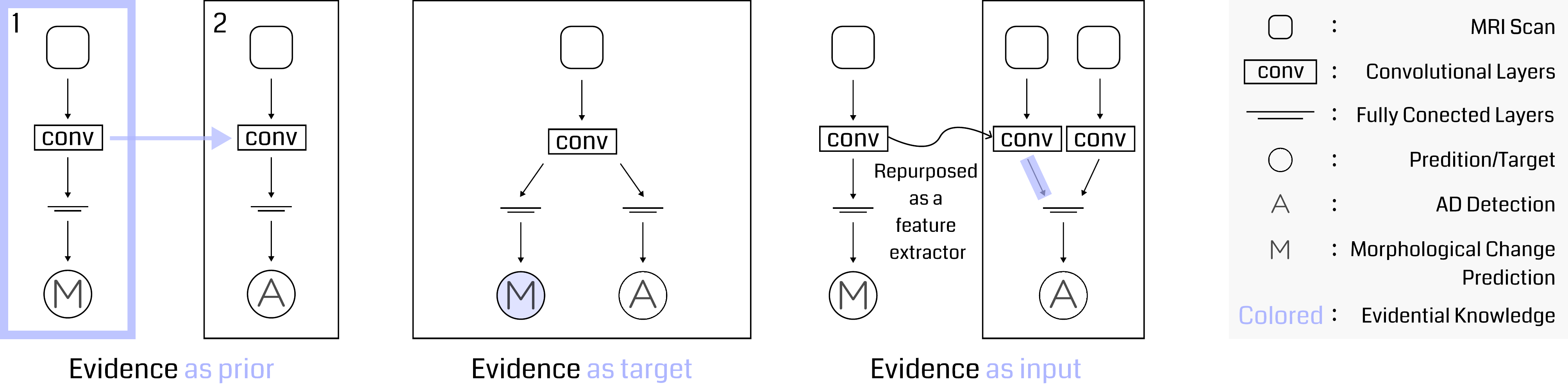}
    \caption{Three different approaches of transfer learning leveraging evidential knowledge as prior, target, and input.}
    \label{fig:2}
\end{figure*}

As a formal framework for transfer learning, we explore the use of evidential knowledge learned from \textit{morphological change prediction} as either model \emph{prior}, \emph{target}, or \emph{input}. Here, we argue that this is ``evidence-empowered'' transfer learning, as it imitates the manual diagnosis of medical consultants which finds clinical evidence from the morphological changes. 
To investigate how evidence-empowered our diagnosis model is, we present a counterfactual inference test by adding random noise to specific brain regions in MRI scans. The counterfactual images ought to yield corresponding changes in diagnosis, but it is not always the case for model-driven diagnosis. 
In experiments on ADNI dataset~\cite{jack2008alzheimer}, we not only empirically show that our approaches consistently outperform baselines regardless of model capacity and data size, but also validate their faithful nature in the counterfactual inference test.

\section{Methodologies}\label{sec:method}

\begin{table*}[ht]\small
    \centering
    \begin{tabular}{llcc|cc|cc||cc}
        \toprule
        &  & \multicolumn{2}{c}{\textbf{3D ResNet-34}} &\multicolumn{2}{c}{\textbf{3D ResNet-50}} &\multicolumn{2}{c}{\textbf{3D ResNet-152}} &\multicolumn{2}{c}{\textbf{Average}} \\

        Method / Auxiliary Task & Input & Acc & AUROC & Acc & AUROC & Acc & AUROC & Acc & AUROC \\
        
        \midrule
        Pre-training / Video Classification & Orig. MRI & 82.6\%& 0.910 & 85.3\% & 0.916 & 84.7\% & 0.917 & 84.2\% & 0.914\\
        Random initialization / None & Orig. MRI & 82.7\%& 0.886  & 84.7\% & 0.903 & 78.4\% & 0.858 & 81.9\% & 0.882\\
        Pre-training / Video Classification & $K = 14$ & 85.1\% & \underbar{0.927}  & 85.8\% & \underbar{0.931} & \underbar{87.6\%} & 0.940 & 86.2\% & \underbar{0.933}\\
        Random initialization / None & $K = 14$ & \underbar{87.4\%} & 0.924 & \underbar{87.5\%} & \underbar{0.931} & 86.5\% & \underbar{0.945} & \underbar{87.1\%} & \underbar{0.933}\\
        
        \midrule
        
        \textbf{EaP}  /  MC Prediction (Ours)& $K = 14$ & 87.5\%& 0.934 & 89.3\% & 0.941 & \textbf{89.4\%} & \textbf{0.949} & 88.7\% & 0.942\\
        \textbf{EaT}  /  MC Prediction (Ours)& $K = 14$ & \textbf{89.2\%} & \textbf{0.944} & 88.9\% &  0.940 & 87.8\% & 0.934 & 88.6\% & 0.939\\
        \textbf{EaI}  /  MC Prediction (Ours)& $K = 14$ & 88.0\% & 0.940 & \textbf{89.9\%} & \textbf{0.943} & 89.2\% & \textbf{0.949} & \textbf{89.0\%} & \textbf{0.944}\\
        
        \bottomrule
        
    \end{tabular}
    \caption{\textbf{Detection Performance:} The top row represents the adopted backbone models; Orig. MRI denotes the original un-segmented MRI scan; Underlines highlight the best baseline for each backbone model and \textbf{Average}; $K=$ \textbf{\#} of parcellations; Note that in this paper, if not specified, the term ``best baseline'' denotes the best \textbf{Average} baseline.}
    \label{tab:1}
\end{table*}

\subsection{Primary Task: Alzheimer's Disease Detection}
\label{ssec:method.1}
Our diagnosis model learns to map a patient case to its corresponding diagnosis result for Alzheimer's disease, \ie, either \texttt{AD} or \texttt{NC} (normal cognition). We formulate the diagnosis of AD as a classification task, namely AD detection. To learn an AD detection model, a patient case's 3D MRI brain image is used as the model input $x_i = \{p_k\}_{k=1}^K$, where $K$ is the number of parcellations. Given a case $x_i$, we adopt 3D-CNN architectures, \eg, 3D ResNet~\cite{he2016deep}, to encode it into a hidden representation $h_i = \text{3D-CNN}(x_i; \theta_{AD})$, where $\theta_{AD}$ represents all the parameters of the 3D-CNN. 
Then, the hidden representation $h_i$ is fed into multiple fully connected layers and ReLU activation functions followed by a softmax non-linear layer predicting the probability distribution $\hat{y}_i$ over classes. 
Given $N$ training samples $(x_i, y_i)$, the parameters $\theta$ of the network are trained to minimize the binary cross-entropy loss $\L_{AD}(\theta)$ of the predicted and true distributions.

\subsection{Auxiliary Task: Morphological Change Prediction}
\label{ssec:aux}
We formulate an auxiliary task, namely morphological change (MC) prediction, where the model learns the knowledge of morphological features by predicting each parcellation's level of atrophy or enlargement. As the morphological features detected in MRI are crucial to AD diagnosis~\cite{asim2018multi, wu2019effects}, this auxiliary task aims to be accomplished with the sole objective of better performing the target primary task.

To collect training data for MC prediction, we acquire 3D MRI scans with clinical information (\eg, age, gender, etc.) from Alzheimer's Disease Neuroimaging Initiative \cite{jack2008alzheimer}. Each image is segmented into 94 parcellations, and we obtain their summary measures (\ie, volume measurements) for $K$ AD-relevant parcellations via FastSurfer~\cite{henschel2020fastsurfer}. A naive setting is using such summary measures as regression labels for MC prediction, but we explore a unified setting for all parcellations, which is the classification of three severity levels of atrophy or enlargement: \texttt{No}, \texttt{Mild}, and \texttt{Severe}. 
More specifically, based on clinical information, we first sort cases into groups. After that, for each group, we use the average volume of NC, mild cognitive impairment (MCI), and AD cases to define the interval for data annotation. The annotation procedure is as follows: If a parcellation's volume is larger than the mean of NC cases' average and MCI cases' average, we label its level of atrophy as \texttt{No}. If a parcellation's volume is smaller than the mean of AD cases' average and MCI cases' average, we label its atrophy level as \texttt{Severe}. Those in between are labeled as \texttt{Mild}. The same logic is applied to the level of enlargement. 

Formally, in morphological change prediction, each parcellation $p_k \in x_i$ is aligned with a classification label $y^k_i$ representing the level of either atrophy or enlargement depending on the $k$-th parcellation type (\ie, brain region). Given a case $x_i$, the model learns to predict $\{y^k_i\}_{k=1}^{K}$. Given $N$ training samples $(x_i, y^{1}_i, y^{2}_i,..., y^{K}_i)$, the model parameters $\theta$ are trained to minimize the aggregated cross-entropy loss $\L_{MC}(\theta)$ between the predicted and ground-truth distributions, \ie, $\L_{MC}(\theta) = \sum_{i=1}^N \sum_{k=1}^{K}  \L^k_{MC}(y^k_i, \hat{y}^k_i;\theta)$.

\subsection{Evidence-Empowered Transfer Learning}\label{method.3}
Figure~\ref{fig:2} illustrates the three different approaches for transfer learning between morphological change prediction and AD detection. Note that, as demonstrated in Section \ref{ssec:aux}, the training data for MC prediction is derived from the data for AD detection without additional manual efforts. Here, not only the knowledge learned from morphological change prediction can be considered AD-specific evidence for AD detection, but this is also a label-efficient way to augment training resources for overcoming the data scarcity problem.

\vspace{1.5mm}
\noindent\textbf{Evidence as Prior (EaP).} Sequential transfer learning has led to promising performance gain~\cite{ebrahimi2020introducing}. The general practice is to pre-train representations on an auxiliary data/task and then adapt these representations to a target primary data/task. 
A common choice of the auxiliary task is image classification with a large scale of natural images such as ImageNet~\cite{russakovsky2015imagenet}. As our target task covers 3D images, as an alternative, we can adopt widely used short-clip video datasets such as Kinetics-700~\cite{smaira2020short} and Moments in Time~\cite{monfortmoments}, where each video corresponds to a single object or event. 
However, as aforementioned, it is sub-optimal to transfer knowledge learned from such non-medical data into AD detection. Our distinction is that we leverage knowledge learned from the AD-relevant task, \ie, MC detection. 
Formally, model parameters $\theta$ are updated by first minimizing the MC prediction loss $\L_{MC}$, and then minimizing the AD detection loss $\L_{AD}$ sequentially:
\begin{align}
    &\text{Pre-training step:}~ \theta_{MC} =  \argmin_{\theta} \L_{MC}(\theta) \\
    &\text{Adaptation step:}~\theta_{AD} =  \argmin_{\theta} \L_{AD}(\theta_{MC} \rightarrow \theta)
\end{align}
where $\rightarrow$ indicates the continual parameter update.

\vspace{1.5mm}
\noindent\textbf{Evidence as Target (EaT).}
The knowledge of morphological features can be used as additional supervision in a multi-task learning (MTL) scheme. 
Here, MC prediction is simultaneously trained with AD detection, with the goal of learning a shared representation that enables the model to consider the morphological features in MRI when performing AD detection. 
The parameters of the model are updated by minimizing the sum of the AD detection and MC prediction losses as:
\begin{align}
   \theta_{AD} =  \argmin_{\theta} \big[ \L_{AD}(\theta) + \lambda \cdot \L_{MC}(\theta) \big]
\end{align}
where $\lambda$ is the preference weight. We empirically set $\lambda$ as 1.

\vspace{1.5mm}
\noindent\textbf{Evidence as Input (EaI).}
Intuitively, the knowledge of morphological features can also be used as additional input features. A straightforward way to do this is to first train an MC prediction model, extract the prediction results for $K$ parcellations, and use them as inputs for the detection model. 
However, since the model predictions are often erroneous\footnote{The performance in MC prediction is 76.0\%, 77.1\%, and 77.6\% in terms of accuracy, when adopting 3D ResNet-34, 50, and 152, respectively.}, we instead adopt the hidden representation from the 3D-CNN encoder for MC prediction as additional input features. That is, the model trained on the MC prediction task is repurposed as a feature extractor for AD detection. 
Once the additional features (\ie, evidential knowledge) are concatenated with the original hidden representation from the 3D-CNN encoder for AD detection as the input of fully-connected layers, two encoders are jointly learned to minimize the AD detection loss.

\begin{figure*}[t]
    \begin{minipage}[t]{1.0\linewidth}
        \centering
        \includegraphics[width=\textwidth]{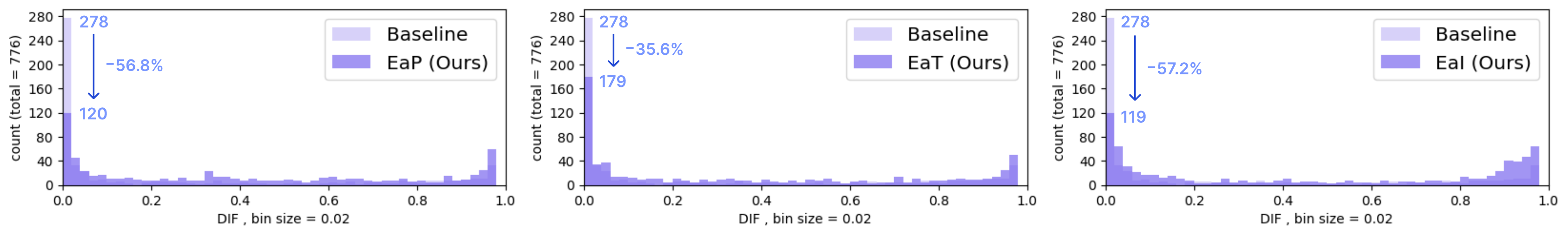}
    \end{minipage}
    \caption{\textbf{Counterfactual Inference Test:} The average result of three adopted backbone models.}
    \label{fig:4}
\end{figure*}

\section{Experiments}
\subsection{Experimental Settings}
\label{ssec:settings}
To evaluate the effectiveness of our three uses of evidential knowledge for transfer learning between MC prediction and AD detection, we apply our approaches and conventional transfer learning to several 3D-CNN architectures. 
Specifically, we adopt 3D ResNet (34, 50, and 152)~\cite{he2016deep} as the backbone models of our framework and baselines, since they have been widely employed for AD detection.
For baselines, we use both (i) randomly initialized weights and (ii) weights pre-trained on two large-scale 3D datasets for recognizing objects and events in videos: Kinetics-700~\cite{smaira2020short} and Moments in Time~\cite{monfortmoments}. We acquired the pre-trained weights from~\cite{hara3dcnns}.

The dataset used in our experiments includes 2781 NC and 1739 AD cases and is split into training, validation, and test sets with a 4:1:1 ratio.
We use accuracy (AD vs. NC) and area under the receiver operating characteristic curve (AUROC) as evaluation metrics. All the experiments in this paper are conducted using a batch size of 16, a learning rate of 1e-5, Adam optimizer~\cite{kingma2014adam}, and OneCycleLR scheduler~\cite{smith2019super}. Experiments are run on one NVIDIA RTX A5000 GPU.

\subsection{Results and Discussions}
\label{sec:results}
We now present the empirical findings of the following three research questions guiding our experiments:

\begin{itemize}
    \item \textbf{RQ1:} \textit{Does our evidence-empowered transfer learning improve AD diagnosis?}
    \item \textbf{RQ2:} \textit{How ``evidence-empowered'' are our models?}
    \item \textbf{RQ3:} \textit{Is our evidence-empowered transfer learning helpful in situations where data is insufficient?}
\end{itemize}

\begin{figure}[t]
    \begin{minipage}[t]{1.0\linewidth}
        \centering
        \includegraphics[width=\textwidth]{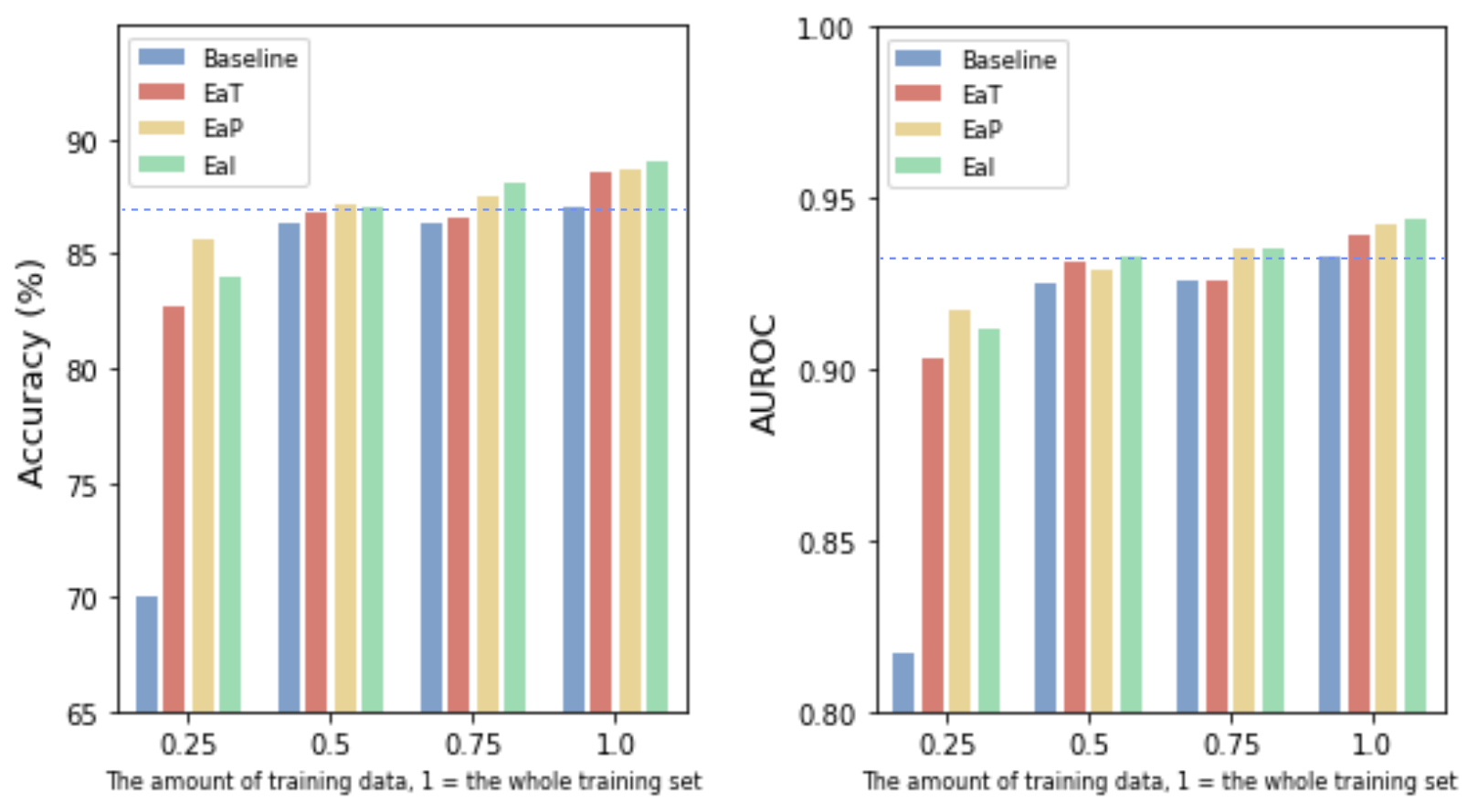}
    \end{minipage}
    \caption{\textbf{Data Efficiency:} The average result of three adopted backbone models. Dotted lines indicate the performance of the best baseline when given 100\% of training data.}
    \label{fig:3}
\end{figure}

\vspace{1mm}
\noindent \textbf{Overall Performance (RQ1). }
Table \ref{tab:1} presents the performance in AD detection of our approaches.
\textbf{EaI} with 3D ResNet-50 achieves the highest accuracy (we empirically set $K = 14$), outperforming the best ResNet-50 baseline by 2.4 percentage points. 
Also, experimental results show that all of our approaches (\textbf{EaP}, \textbf{EaT}, and \textbf{EaI}) result in higher detection accuracy than both randomly initialized and pre-trained baselines, regardless of model capacity.
Moreover, comparing baseline performances, we find that despite its massive pre-training data, baselines pre-trained with video classification tasks have lower accuracy than not only our approaches but also randomly initialized baselines (when $K = 14$). Similar findings of such negative transfer are also reported in~\cite{kora2021transfer, pan2009survey}. This demonstrates that our evidence-empowered transfer learning framework improves AD diagnosis more effectively than the conventional transfer learning approach.

\vspace{1.5mm}
\noindent \textbf{Faithfulness (RQ2). }
In this section, we investigate how ``evidence-empowered'' the proposed framework is. That is, we look into if the diagnosis result is faithful to changes in evidence (\ie, changes in MRI scans).
For that purpose, we present a counterfactual inference test, where we deliberately add random noise to parcellations whose morphological changes are annotated as \texttt{Severe} (\ie, \textit{corrupt the evidence}) to acquire counterfactual samples. 
After that, we measure the difference $DIF$ between predictions when the model is given a pair of an original sample $x$ and its counterfactual $\bar{x}$ (776 pairs in total). Let $P(x)$ and $P(\bar{x})$ be the outputted softmax probabilities (AD) when given $x$ and $\bar{x}$. The difference $DIF$ is calculated as follow: $DIF = |P(\bar{x}) - P(x)|$.

Results of the counterfactual inference test are shown in Figure~\ref{fig:4}. We can see that all of our approaches have lower counts in the first bins ($0 \leq DIF < 0.02 $) compared to the best baseline. 
This means that when given a pair of an original sample and its counterfactual, our models can output predictions that better reflect the changes in MRI scans. In other words, our models have more sample pairs yielding larger $DIF$ (from the second to the last bin). 
This pattern not only allows us to quantify how ``evidence-empowered'' our framework is, but also suggests that our approaches are faithful to changes in evidence and reflect these changes in their diagnosis better than model-driven approaches do.

\vspace{1.5mm}
\noindent \textbf{Data Efficiency (RQ3). }
We further experiment on varying amounts of training data as a stress test for data-scarce scenarios. The results are presented in Figure~\ref{fig:3}.
Firstly, we find that our approaches always outperform the best baseline when the same amount of training data is given. 
Furthermore, when only 25\% of the training data is given, our approaches (\textbf{EaT}, \textbf{EaI}, and \textbf{EaP}) outperform the best baseline by 12.7, 14.0, and 15.7 percentage points in terms of accuracy, respectively.
Secondly, comparing to the best baseline trained with 100\% of training data, \textbf{EaP} and \textbf{EaI} both achieve (i) higher accuracy with only 50\% and 75\% (respectively) of training data and (ii) higher AUROC with 75\% of training data, demonstrating that our approaches can show comparable performance even when only a limited amount of training data is available (\ie, data efficiency).
These experimental findings affirm the data efficiency of our framework, which is an important property as AD diagnosis is often limited by the data scarcity problem.

\section{Conclusions}
\label{sec:conclusion}
This paper presents an evidence-empowered transfer learning with an AD-relevant auxiliary task named \textit{morphological change prediction}. With the evidential knowledge learned from this auxiliary task, we explore the use of evidence as model \emph{prior}, \emph{target}, or \emph{input}. Applying our framework to AD diagnosis, our models not only outperform baselines regardless of model capacity and data size, but also manifest their faithful nature in the counterfactual inference test.
In the future, as our framework is label-efficient, we presume our framework can be adapted to diverse medical imaging fields to mitigate the data scarcity problem.

\section{Acknowledgments}
\label{sec:acknowledgments}
This work is supported by Samsung Research Funding Center of Samsung Electronics (Project Number SRFC-TF2103-01). 

\section{Compliance with Ethical Standards}
Ethical approval is not required (https://adni.loni.usc.edu).
\bibliographystyle{IEEEbib}
\bibliography{strings, refs}

\begin{thebibliography}{10}

\bibitem{setio2017validation}
Arnaud Arindra~Adiyoso Setio, Alberto Traverso, Thomas De~Bel, Moira~SN Berens,
  Cas Van Den~Bogaard, Piergiorgio Cerello, Hao Chen, Qi~Dou, Maria~Evelina
  Fantacci, Bram Geurts, et~al.,
\newblock ``Validation, comparison, and combination of algorithms for automatic
  detection of pulmonary nodules in computed tomography images: the luna16
  challenge,''
\newblock {\em Medical image analysis}, vol. 42, pp. 1--13, 2017.

\bibitem{simpson2019large}
Amber~L Simpson, Michela Antonelli, Spyridon Bakas, Michel Bilello, Keyvan
  Farahani, Bram Van~Ginneken, Annette Kopp-Schneider, Bennett~A Landman, Geert
  Litjens, Bjoern Menze, et~al.,
\newblock ``A large annotated medical image dataset for the development and
  evaluation of segmentation algorithms,''
\newblock {\em arXiv preprint arXiv:1902.09063}, 2019.

\bibitem{kruthika2019multistage}
KR~Kruthika, HD~Maheshappa, Alzheimer's Disease~Neuroimaging Initiative,
  et~al.,
\newblock ``Multistage classifier-based approach for alzheimer's disease
  prediction and retrieval,''
\newblock {\em Informatics in Medicine Unlocked}, vol. 14, pp. 34--42, 2019.

\bibitem{jack2008alzheimer}
Clifford~R Jack~Jr, Matt~A Bernstein, Nick~C Fox, Paul Thompson, Gene
  Alexander, Danielle Harvey, Bret Borowski, Paula~J Britson, Jennifer
  L.~Whitwell, Chadwick Ward, et~al.,
\newblock ``The alzheimer's disease neuroimaging initiative (adni): Mri
  methods,''
\newblock {\em Journal of Magnetic Resonance Imaging: An Official Journal of
  the International Society for Magnetic Resonance in Medicine}, vol. 27, no.
  4, pp. 685--691, 2008.

\bibitem{he2016deep}
Kaiming He, Xiangyu Zhang, Shaoqing Ren, and Jian Sun,
\newblock ``Deep residual learning for image recognition,''
\newblock in {\em Proceedings of the IEEE conference on computer vision and
  pattern recognition}, 2016, pp. 770--778.

\bibitem{asim2018multi}
Yousra Asim, Basit Raza, Ahmad~Kamran Malik, Saima Rathore, Lal Hussain, and
  Mohammad~Aksam Iftikhar,
\newblock ``A multi-modal, multi-atlas-based approach for alzheimer detection
  via machine learning,''
\newblock {\em International Journal of Imaging Systems and Technology}, vol.
  28, no. 2, pp. 113--123, 2018.

\bibitem{wu2019effects}
Zhanxiong Wu, Dong Xu, Thomas Potter, Yingchun Zhang, and Alzheimer's
  Disease~Neuroimaging Initiative,
\newblock ``Effects of brain parcellation on the characterization of
  topological deterioration in alzheimer's disease,''
\newblock {\em Frontiers in aging neuroscience}, vol. 11, pp. 113, 2019.

\bibitem{henschel2020fastsurfer}
Leonie Henschel, Sailesh Conjeti, Santiago Estrada, Kersten Diers, Bruce
  Fischl, and Martin Reuter,
\newblock ``Fastsurfer-a fast and accurate deep learning based neuroimaging
  pipeline,''
\newblock {\em NeuroImage}, vol. 219, pp. 117012, 2020.

\bibitem{ebrahimi2020introducing}
Amir Ebrahimi, Suhuai Luo, and Raymond Chiong,
\newblock ``Introducing transfer learning to 3d resnet-18 for alzheimer’s
  disease detection on mri images,''
\newblock in {\em 2020 35th international conference on image and vision
  computing New Zealand (IVCNZ)}. IEEE, 2020, pp. 1--6.

\bibitem{russakovsky2015imagenet}
Olga Russakovsky, Jia Deng, Hao Su, Jonathan Krause, Sanjeev Satheesh, Sean Ma,
  Zhiheng Huang, Andrej Karpathy, Aditya Khosla, Michael Bernstein, et~al.,
\newblock ``Imagenet large scale visual recognition challenge,''
\newblock {\em International journal of computer vision}, vol. 115, no. 3, pp.
  211--252, 2015.

\bibitem{smaira2020short}
Lucas Smaira, Jo{\~a}o Carreira, Eric Noland, Ellen Clancy, Amy Wu, and Andrew
  Zisserman,
\newblock ``A short note on the kinetics-700-2020 human action dataset,''
\newblock {\em arXiv preprint arXiv:2010.10864}, 2020.

\bibitem{monfortmoments}
Mathew Monfort, Alex Andonian, Bolei Zhou, Kandan Ramakrishnan, Sarah~Adel
  Bargal, Tom Yan, Lisa Brown, Quanfu Fan, Dan Gutfruend, Carl Vondrick,
  et~al.,
\newblock ``Moments in time dataset: one million videos for event
  understanding,''
\newblock {\em IEEE Transactions on Pattern Analysis and Machine Intelligence},
  pp. 1--8, 2019.

\bibitem{hara3dcnns}
Kensho Hara, Hirokatsu Kataoka, and Yutaka Satoh,
\newblock ``Can spatiotemporal 3d cnns retrace the history of 2d cnns and
  imagenet?,''
\newblock in {\em Proceedings of the IEEE Conference on Computer Vision and
  Pattern Recognition (CVPR)}, 2018, pp. 6546--6555.

\bibitem{kingma2014adam}
Diederik~P Kingma and Jimmy Ba,
\newblock ``Adam: A method for stochastic optimization,''
\newblock {\em arXiv preprint arXiv:1412.6980}, 2014.

\bibitem{smith2019super}
Leslie~N Smith and Nicholay Topin,
\newblock ``Super-convergence: Very fast training of neural networks using
  large learning rates,''
\newblock in {\em Artificial intelligence and machine learning for multi-domain
  operations applications}. SPIE, 2019, vol. 11006, pp. 369--386.

\bibitem{kora2021transfer}
Padmavathi Kora, Chui~Ping Ooi, Oliver Faust, U~Raghavendra, Anjan Gudigar,
  Wai~Yee Chan, K~Meenakshi, K~Swaraja, Pawel Plawiak, and U~Rajendra Acharya,
\newblock ``Transfer learning techniques for medical image analysis: A
  review,''
\newblock {\em Biocybernetics and Biomedical Engineering}, 2021.

\bibitem{pan2009survey}
Sinno~Jialin Pan and Qiang Yang,
\newblock ``A survey on transfer learning,''
\newblock {\em IEEE Transactions on knowledge and data engineering}, vol. 22,
  no. 10, pp. 1345--1359, 2009.

\end{thebibliography}

\end{document}